\documentstyle[preprint,epsfig,aps]{revtex}
\tightenlines

\newcommand{\bnl}           {$\rm^{1}$}
\newcommand{\ires}          {$\rm^{2}$}
\newcommand{\kraknuc}       {$\rm^{3}$}
\newcommand{\krakow}        {$\rm^{4}$}
\newcommand{\baltimore}     {$\rm^{5}$}
\newcommand{\newyork}       {$\rm^{6}$}
\newcommand{\nbi}           {$\rm^{7}$}
\newcommand{\texas}         {$\rm^{8}$}
\newcommand{\bergen}        {$\rm^{9}$}
\newcommand{\bucharest}     {$\rm^{10}$}
\newcommand{\kansas}        {$\rm^{11}$} 
\newcommand{\oslo}          {$\rm^{12}$}
\begin{document}

\title{ Rapidity dependence of antiproton to proton ratios in Au+Au collisions at $\sqrt{s_{NN}}=130$ GeV}

\author{
  I.~G.~Bearden\nbi, 
  D.~Beavis\bnl, 
  C.~Besliu\bucharest,
  Y.~Blyakhman\newyork,
  J.~Brzychczyk\krakow,
  B.~Budick\newyork,
  H.~B{\o}ggild\nbi,
  C.~Chasman\bnl,
  C.~H.~Christensen\nbi,
  P.~Christiansen\nbi,
  J.~Cibor\kraknuc,
  R.~Debbe\bnl,
  J.~J.~Gaardh{\o}je\nbi,
  K.~Grotowski\krakow,
  K.~Hagel\texas,
  O.~Hansen\nbi,
  A.~Holm\nbi,
  A.~K.~Holme\oslo,
  H.~Ito\kansas,
  E.~Jakobsen\nbi,
  A.~Jipa\bucharest,
  J.~I.~J{\o}rdre\bergen,
  F.~Jundt\ires,
  C.~E.~J{\o}rgensen\nbi,
  T.~Keutgen\texas,
  E.~J.~Kim\baltimore,
  T.~Kozik\krakow,
  T.~M.~Larsen\oslo,
  J.~H.~Lee\bnl,
  Y.~K.~Lee\baltimore,
  G.~L{\o}vh{\o}iden\oslo,
  Z.~Majka\krakow,
  A.~Makeev\texas,
  B.~McBreen\bnl,
  M.~Murray\texas,
  J.~Natowitz\texas,
  B.~S.~Nielsen\nbi,
  K.~Olchanski\bnl,
  J.~Olness\bnl,
  D.~Ouerdane\nbi,
  R.~P\l aneta\krakow,
  F.~Rami\ires,
  D.~R{\"o}hrich\bergen,
  B.~H.~Samset\oslo,
  S.~J.~Sanders\kansas,
  R.~A.~Sheetz\bnl,
  Z.~Sosin\krakow,
  P.~Staszel\nbi,
  T.~F.~Thorsteinsen\bergen$^+$,
  T.~S.~Tveter\oslo,
  F.~Videb{\ae}k\bnl,
  R.~Wada\texas,
  A.~Wieloch\krakow, and 
  I.~S.~Zgura\bucharest\\
  (BRAHMS Collaboration)\\[1ex]
  \bnl~Brookhaven National Laboratory, Upton, New York 11973,\\ 
  \ires~Institut de Recherches Subatomiques and Universit{\'e} Louis Pasteur, Strasbourg, France,\\
  \kraknuc~Institute of Nuclear Physics, Krakow, Poland,\\
  \krakow~Jagiellonian University, Krakow, Poland,\\
  \baltimore~Johns Hopkins University, Baltimore, Maryland 21218,\\
  \newyork~New York University, New York, New York 10003,\\
  \nbi~Niels Bohr Institute, University of Copenhagen, Denmark,\\
  \texas~Texas A$\&$M University, College Station, Texas 77843,\\
  \bergen~University of Bergen, Department of Physics, Bergen, Norway,\\
  \bucharest~University of Bucharest, Romania,\\
  \kansas~University of Kansas, Lawerence, Kansas 66045,\\
  \oslo~University of Oslo, Department of Physics, Oslo, Norway,\\
  $^+ Deceased$}

\date{April 30, 2001}

\maketitle

\begin{abstract}
  Measurements, with the BRAHMS detector, of the antiproton to proton
  ratio at central and forward rapidities are presented for Au+Au reactions at
  $\sqrt{s_{NN}}=130$ GeV, and for three different
  collision centralities. For collisions in the $0-40\%$ centrality range
  we find $N(\bar{{\rm p}})/N({\rm p}) = 0.64 \pm 0.04_{(stat.)} \pm 0.06_{(syst.)} $ at 
$y \approx 0$, $0.66 \pm 0.03 \pm 0.06$ at $y\approx 0.7$, and $0.41 \pm
  0.04 \pm 0.06$ at $y \approx 2$. The ratios are found to be nearly
  independent of collision centrality and transverse momentum. 
  The   measurements demonstrate that the antiproton and proton rapidity
  densities vary differently with rapidity, and indicate that a
  net-baryon free midrapidity plateau (Bjorken limit) is not reached
  at this RHIC energy.  \\[1ex]
  PACS numbers: 25.75.-q 
\end{abstract}

The reaction mechanism between heavy ions at high energies is expected
to evolve from full stopping to complete transparency with increasing
collision energy. In the case of full stopping, the baryons of the
colliding nuclei will be shifted from the rapidity of the incident
beam to midrapidity ($y\approx 0$), leading to the formation of a
central zone with significant net-baryon density. In the case of full
transparency, also called the Bjorken limit~\cite{Bjorken83}, the
baryons from the interacting nuclei will, after the collision, also
be shifted  from beam rapidity, but mid-rapidity will be devoid of original baryons. 
In this region, the
net-baryon density is zero, but the energy density is high. Almost complete
stopping is observed for Au+Au reactions at AGS energies
($\sqrt{s_{NN}} \approx 5~$GeV). 
In reactions between lead nuclei
at SPS energies ($\sqrt{s_{NN}}=17$~GeV), transparency begins to
set in, and systematics suggest that maximum baryon density occurs at
energies intermediate between AGS and SPS (see
e.g. \cite{Her99,Vid95}). The situations of maximum baryon density and
of vanishing net-baryon density at midrapidity give rise to entirely
different initial conditions for the possible creation of a deconfined
quark-gluon system. 

The rapidity dependence of the antiproton to proton ratio in
collisions between Au nuclei at the Relativistic Heavy Ion Collider
(RHIC), Brookhaven National Laboratory, is investigated for
$\sqrt{s_{NN}}=130$ GeV, the highest center of mass energy yet
achieved in collisions between heavy nuclei in the laboratory. The
data were collected with the BRAHMS detector during the final 2 weeks
of the first RHIC run in August and September 2000, where the beam
luminosity reached $\approx 10\%$ of the nominal design value 
($2\times10^{26}{\rm cm}^{-2}{\rm s}^{-1}$). We present
measurements of the $N(\bar{{\rm p}})/N({\rm p})$  rapidities $y\approx 0$,
$0.7$ and $2$ as a function of collision centrality and transverse
momentum together with $N(\pi^{-})/N(\pi^{+})$ ratios at $y\approx 0,
1$ and $3$. The beam rapidity is 4.95.  The measurements provide the
first particle ratios over an extended rapidity range at RHIC
energies. 
We find that while the pion ratios are close to unity,
the measured antiproton to proton ratio decreases from $0.64 \pm
0.04_{(stat.)} \pm 0.06_{(syst.)}$ at 
$y\approx~0$ and $0.66 \pm 0.03_{(stat.)} \pm 0.06_{(syst.)}$ at $y\approx~0.7$ to
$0.41 \pm 0.04_{(stat.)} \pm0.06_{(syst.)}$ at $y\approx~2$ for 0-40\% collision
centrality. The results show that a net-baryon free midrapidity region
has not been attained, although the $N(\bar{{\rm p}})/N({\rm p})$ ratio is the
highest that has been measured so far in nucleus-nucleus
collisions. Recently, the STAR collaboration has measured similar
particle ratios at $y\approx~0$ in a narrower momentum range
\cite{STAR01} and their results agree well with the $y\approx~0$
measurements presented here.

The BRAHMS detector system~\cite{BRAHMSNIM} used in the present
measurements consists of two independent magnetic spectrometers that
can be positioned over an
angular range from $2.3^\circ$ to $30^\circ$
(forward spectrometer, FS) and from $25^\circ$ to $90^\circ$
(midrapidity spectrometer, MRS) with respect to the beam line.
A scintillator tile multiplicity array (TMA) measures charged particle
emission in the central pseudorapidity region ($-2.0<\eta <+2.0$) and
is used for off--line centrality selection.  This detector consists of
38 square tiles of plastic scintillators ($12 \times 12 \times 0.5
\;$cm$^{3}$). The tiles are grouped in rows of 8 to form a tube of
hexagonal cross section with the axis along the beam pipe, such that
no tile is placed in the MRS and FS acceptances.
Two global detector systems cover forward angles.
The Zero Degree Calorimeters (ZDC) at $\pm 18$ m measure
spectator neutrons~\cite{ZDCNIM}. The Beam-Beam Counters
(BB) consist of two arrays with a total of 70 phototubes, each with a
Cerenkov radiator, positioned $\pm$~2.15 m and measure charged hadrons in the pseudo-rapidity
range of $3.0<|\eta|<3.8$.  
These two systems are used to define collision events by two simultaneous 
measurements of the interaction vertex position.

For each actual vertex position 
the total energy deposited in each ring of TMA tiles is
determined. This energy loss signal is transformed to a charged
particle multiplicity via division with the expected average energy
deposited by one primary charged particle in a tile at its corresponding
pseudo-rapidity.  
Centrality cuts are
applied by selecting appropriate ranges in the multiplicity
spectrum. The cuts can be expressed in terms of the fraction of the
nuclear reaction cross section by normalization to the integral of the
TMA spectrum obtained with a minimum bias trigger. This trigger
requires energy deposition in each of the two  ZDCs above 
$25$ GeV with the additional condition of at least one
tile having a hit. 
Comparing to simulations with the HIJING code
this requirement selects events corresponding to
$\approx 99 \pm 2 \%$ of the nuclear interaction cross section.  Centrality
bins from 0 to 10~\%, 10 to 20~\%, and 20 to 40~\%, of this event selection
were used in the analysis.

In the present measurements the MRS was operated at both $90^\circ$
and $40^\circ$
and the FS at $4^\circ$.
The magnets of the two spectrometers were operated at fields
allowing the reconstruction of particle tracks with momenta above
$\approx 0.2$ GeV$/c$ in the MRS and above $\approx 2$ GeV$/c$ in
the FS. The solid angles subtended by the MRS and FS are 6.5 msr and 
0.8 msr, respectively. Each spectrometer consists of two Time
Projection Chambers (TPCs) positioned on either side of a dipole
magnet and followed by a segmented scintillator time of flight (TOF)
wall for particle identification (PID). In the case of the FS this
tracking and PID arrangement is preceded by an additional dipole
magnet to sweep particles away from the beam and reduce background. 

Particle momenta are determined by projecting the straight line tracks
as reconstructed in the two TPCs to the magnet and calculating the
bending angle of matched tracks using an effective edge
approximation. The momentum resolution is $\delta p/p \approx 0.01
\cdot p$ for the FS, and for the MRS $\approx 0.04 \cdot
p~(90^{\circ})$, and $\approx0.03\cdot p~(40^{\circ})$ at the field settings
used.  MRS tracks are required to originate from the primary
vertex as determined by the BB counters within $\pm 5$ cm horizontally,
and $\pm 5$ cm verticaly from the nominal beam position.  Tracks in the FS have the same vertical,
but a much loser
horizontal matching requirement ($\pm 12$ cm) due to the uncertainty
associated with projecting tracks back to the beam line.  
PID requires not only a TOF measurement, but also the determination of the flight distance and
thus of the collision vertex position on an event by event basis.
The BB counters have an
intrinsic time resolution of 65 ps and permit a determination of the
collision vertex position to a precision of $\approx 2$ cm by
measuring the difference in arrival time of particles in the
two arrays, assuming that particles travel with the
velocity of light.  This method is confirmed by determining the vertex
by projecting tracks determined in the first TPC of the MRS back to
the beam plane.  The collision vertex distribution was approximately of
Gaussian shape with $\sigma \approx 70$ cm. 
We select events with vertex between $\pm 15$ cm for MRS and $\pm 40$ cm for FS, respectively.
The two TOF arrays are positioned at 4.3 m (MRS) and
8.6 m (FS) from the nominal interaction vertex position, respectively.
The overall time resolution was determined to be $\sigma$(TOF) $
\approx 120$ ps. 

\begin{figure}[htp]
  \epsfig{file=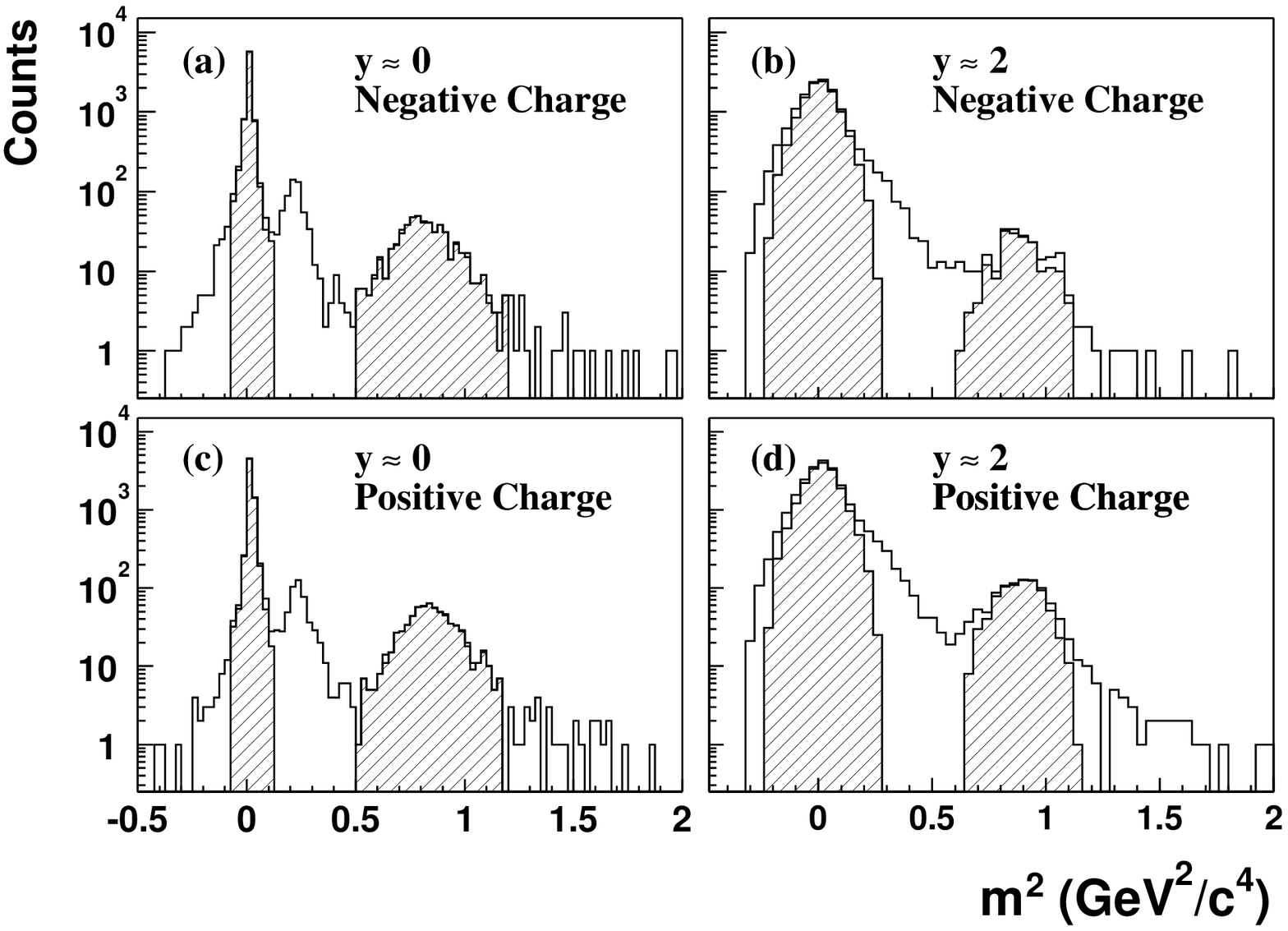,width=7.5cm}
  \caption{ 
    Distributions of $m^{2}$ for charged particles identified in the BRAHMS
    spectrometers. Left panels (a) and (c) show data  
    from $90^\circ$
    for negatively and positively
    charged particles, respectively.  The hatched areas show the pions
    selected for analysis in the momentum range $p$ $< 1.6$ GeV/$c$ and
    $-0.075<m^2<0.125$ and the protons for $ p <  2.4$ GeV/$c$ and
    $0.5<m^2<1.2$.
    The right panels (b) and (d) show data  from $4^\circ$.
    The hatched region for these panels shows yield
    of particles selected for analysis on the basis of fiducial cuts
    on time--of--flight for a momentum range of $2<p<4$ GeV/$c$.}
  \label{fig1}
\end{figure}

Kaons and protons are seperated
in the momentum range $p<2.4~\rm{GeV}$$/c$ and $p<4.5~ \rm{GeV}$$/c$
in the MRS and FS, respectively.
Kaons and pions can be separated up to $p= 1.6$
GeV/$c$ in MRS. In the FS, kaons could not be cleanly separated from pions,
although the $N(\pi^{-})/N(\pi^{+})$ ratio for $0.15<p_{t}<0.3$
GeV/$c$ was determined with only a small kaon contamination.
Figure~\ref{fig1} demonstrates the PID achieved in the MRS and FS. The
lower two panels show the $m^2$ spectra obtained for positively
charged particles ($\pi^+, K^+, {\rm p}$) and the upper two panels for
negative particles ($\pi^-, K^-, {\bar{\rm p}}$). These distributions were
calculated using $m^2 = p^2 (t^2/L^2-1)$, where $p$, $t$, and $L$
denote the particle momentum, TOF, and flight distance,
respectively. The MRS data shown are from $0.4<p_{t}<2.4$ GeV/$c$,
while the FS data are from $0.15<p_{t}<0.55$ GeV/$c$.

The particle yield in the FS is determined by selecting tracks having
a TOF within a $\pm 2\sigma$ band of the expected TOF vs. momentum for
a given particle type. The results of such cuts for pions and protons
are shown in Fig.~\ref{fig1}.
There is a systematic uncertainty that causes a small shift in the
$m^2$ scale vs momentum. The effect on extracted particle ratios from the prescribed method is estimated to be small. 
%The separation in $m^{2}$ get increasingly better at lower momenta.  
In the MRS, where the momenta are much lower and the particle peaks better separated,
the yields can be  determined by applying cuts in the $m^2$ spectra,
as seen in Fig.~\ref{fig1}.
The number of particles measured at either
polarity is normalized to the number of collision events
defined by the BB--ZDC coincidences fulfilling a given centrality cut
described above and the ratios are calculated. 

The acceptances for the spectrometers for positively charged particles at one field are
equal to the acceptances for negatively charge particles at the opposite polarity.
Thus in ratios of numbers of particles, measured at opposite field polarities
most systematic errors cancel out. 
The ratios have been corrected for losses of
antiprotons due to annihilation evaluated by GEANT
simulations to be less than  2~\% in the MRS  and about 3.5~\% in the FS. 
It is noted that background from misidentified tracks and from the tails
of kaon and pion peaks are negligible for the
measurements at $90^\circ$ and $40^\circ$.
At forward angles the contributions from background tracks
increase but are still small, and affect the extracted ratios negligibly ($<1\%$). 
In the MRS, the background contribution from slow protons, arising mainly
from the interaction of pions with the Be beam pipe, is $\approx10\%$
for the lowest $p_{t}$ bin.  The data have been corrected for this effect.
In the FS this contribution is found to be negligible.
The main sources of systematic uncertainties in the particle ratios
are due to normalizations, corrections for the background
of slow protons, and from corrections for antiproton absorption. The
normalizations contribute less than $\pm$5\%. 

\begin{figure}[htp]
 \epsfig{file=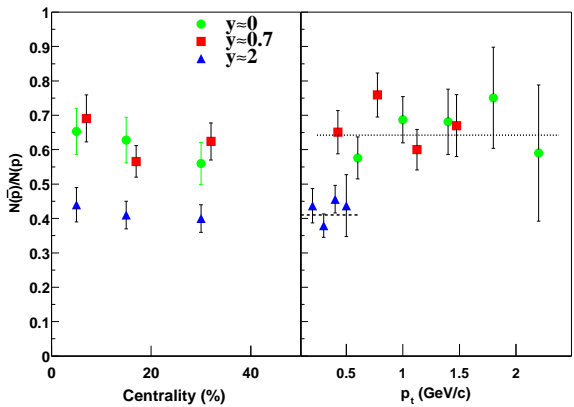,width=7.5cm}
  \caption{
    The left panel shows the centrality dependence of the $N(\bar{{\rm p}})/N({\rm p})$ ratios for
    the three  rapidity values: $y \approx~0$ (filled circles), $y
    \approx 0.7$ (open squares) and $y \approx
    2$ (filled triangles). Only statistical errors are shown. 
The data points for $y \approx~0.7$ are shifted 
slightly for display purpose. 
    The centrality percentages are described in the text. The right panel shows the 
    transverse momentum dependence
    of the measured $N(\bar{{\rm p}})/N({\rm p})$ ratio for the same three rapidity
    intervals for events selected from the $0-40\%$ centrality cut. The upper dotted line shows the
    average ratio for $y \approx~0$, while the dashed line for $y \approx~2$.
 }
  \label{fig2}
\end{figure}

Figure~\ref{fig2} shows the dependence of the measured
$N(\bar{{\rm p}})/N({\rm p})$ ratios on collision centrality and on particle
transverse momentum. Ratios were corrected for antiproton absorption
as described above.  Figure~\ref{fig2}(a) shows the centrality dependence
for data summed over all momentum bins, while  Fig.~\ref{fig2}b shows
the $p_{t}$ dependence for data summed over all centrality bins. The
vertical error bars represent statistical errors  only.  
It is seen that the centrality dependence of the ratio is small for the three considered
rapidities. 
The $N(\pi^{-})/N(\pi^{+})$  ratios (not shown) exibit a similar lack of
centrality and $p_{t}$ dependence.   

The ratios shown in Fig.~\ref{fig2} have not been
corrected for protons and antiprotons that originate from weak
decays of hyperons ($\Lambda, \Sigma$, $etc$). 
The correction factors will in general depend on
the relative  production ratios of hyperons ($N(H$)) 
and primary baryons ($N(B$)) and
their antiparticles, and on the relative slopes of their
spectra. We have studied the  magnitude of the corrections using
various assumptions as input to  GEANT simulations.
Assuming primary $N(H$)/$N(B$) ratios of up to $0.5$ we find that
the correction to the quoted ratios is less than $\pm5\%$ for
 $N({\bar H})/N(H)$  between 0.4 and 0.8 (y~$\approx~0$) and  
between 0.3 and 0.5 (y~$\approx$2). 

\begin{figure}[htp]
  \epsfig{file=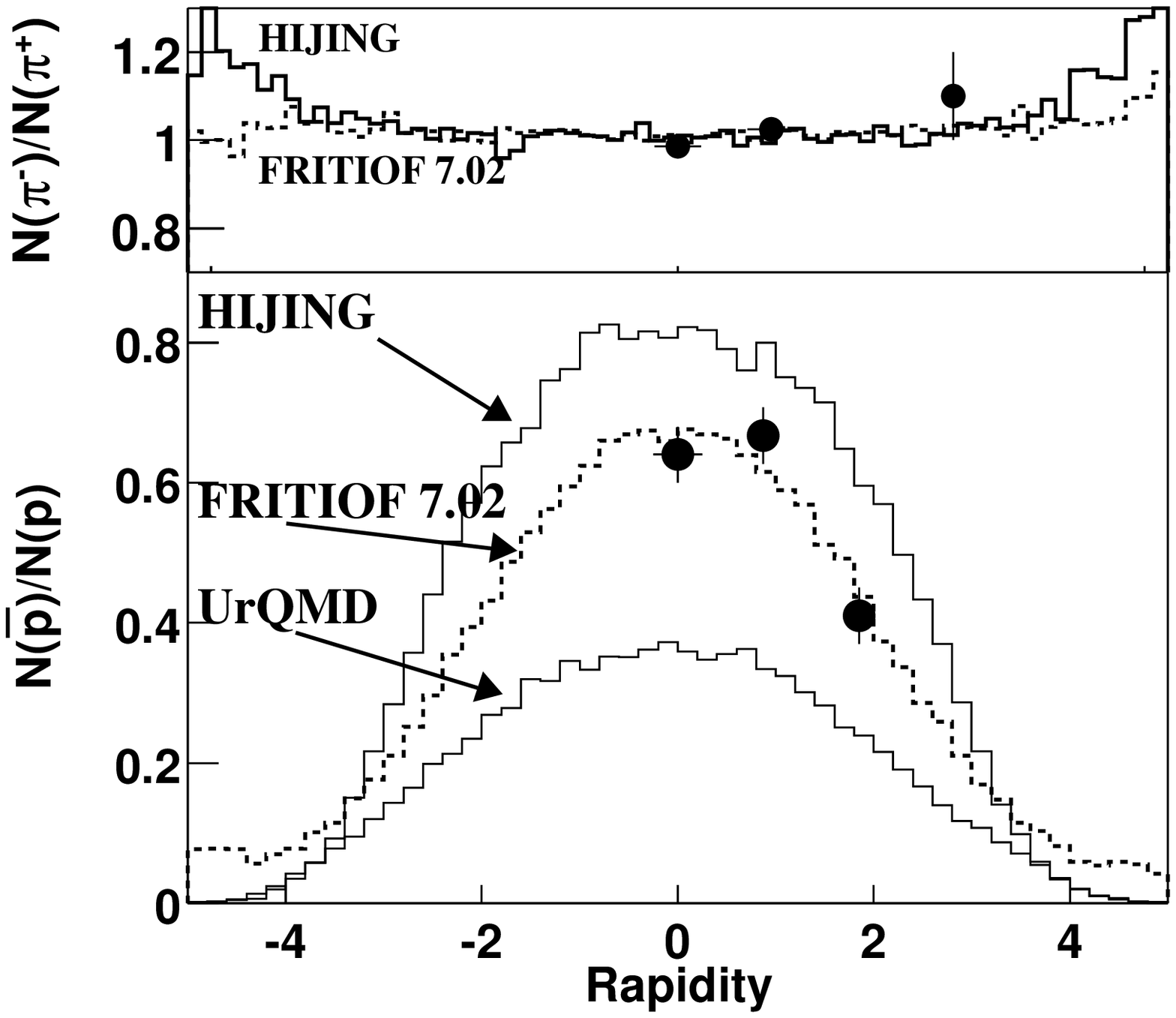,width=7.5cm}
  \caption{
    Comparison of the measured  $N(\bar{{\rm p}})/N({\rm p})$ (lower panel) and
    $N({\pi}^-)/N({\pi}^+)$ (upper panel) ratios to model
    predictions. The data shown are for $0-40\%$ central events and
integrated over the transverse momentum  range shown in Fig.~\ref{fig2}. 
    The three model calculations (HIJING,
    FRITIOF, and UrQMD) are shown for comparison. See text for details.} 
  \label{fig3}
\end{figure}

Figure~\ref{fig3} summarizes the rapidity dependence of the measured
particle ratios  calculated for the $0-40 \%$ most central
collisions. The upper and lower panel depict the $N(\pi^{-})/N(\pi^{+})$  and
$N(\bar{{\rm p}})/N({\rm p})$ ratios, respectively. 
It is seen that, while the pion ratio is
independent of rapidity and consistent with unity, the antiproton to
proton ratio drops significantly with increasing rapidity.   

The $N(\bar{{\rm p}})/N({\rm p})$ ratio of 0.64 found here at $y\approx~0$ is
considerably higher than the similar ratios measured in Pb+Pb
collisions at the SPS ($\approx 0.07-0.15$ at $\sqrt{s_{NN}}$ = 17 GeV)
\cite{SPSppbar,Sick99} and at the AGS ($\approx 2.5\cdot10^{-4}$ at
$\sqrt{s_{NN}}=5$ GeV) \cite{Ahle98}. It is, in fact, close to the p+p
result ($0.61\pm 0.10$) from the ISR \cite{Guettler76} at
$\sqrt{s_{NN}}=63$ GeV and $p_t \approx 0.3 $GeV/$c$ , and below the
value obtained by extrapolating the p+p systematics to 130
GeV CM energy.  The rapidity dependence of the ratios is of
particular interest.  The $N(\bar{{\rm p}})/N({\rm p})$ at $y\approx~0$ and
$\approx~0.7$ are approximately the same, consistent with the formation
of a plateau around mid-rapidity.
The decrease in the ratio over the next unit
of rapidity is larger than observed in A+A collisions at lower
energies, but are very similar to the p+p result at roughly half the
CM energy \cite{Capi74}.

Finally, in Fig.~\ref{fig3}, we compare the measured ratios to
calculations using the HIJING model ~\cite{HIJING}, the FRITIOF
7.02 string model~\cite{FRITIOF}  and the UrQMD cascade model
~\cite{UrQMD} using the same centrality cuts as in the data
analysis. Hyperon decays have not been included in the calculations shown, 
but affect the results by less than $5~\%$.  
All three models reproduce the observed pion ratios well.
FRITIOF reproduces our $N(\bar{{\rm p}})/N({\rm p})$ ratios, while
overpredicting (by $\approx 30~\%$) the charged particle yield at $\eta
\approx 0$ ~\cite{Phobos00}.  This is due to a significant
degree of stopping in the model.  On the other hand,
HIJING which describes the overall charged particle yields at
$\eta\approx0$ fails in describing the antiproton to proton ratio. 
This feature of the model is related to the small stopping of
the projectile baryons.
The UrQMD model, which is not a partonic model, under predicts the ratio by nearly a factor of two.
These models exemplify the present theoretical understanding of heavy
ion collisions in this new energy regime. None of the models offer a
consistent description of the observed features.

In summary, the BRAHMS experiment has measured the ratio of positive
and negative pions and protons at central and forward rapidities. We
find that the pion ratios are close to unity as would be expected at
these energies, where about $4000$ charged particles (predominantly 
pions) are produced per central collision.
We find, however, that for central
collisions at  $\sqrt{s_{NN}} = 130$ GeV the ratio of antiprotons to
protons is still  significantly below unity at midrapidity and
decreases towards forward rapidity. 
The rapidity dependence serves as
an indicator of the balance between baryon number transport to the 
central region and antibaryon and baryon pair production. 
This suggests that there is still a significant contribution from participant baryons over the
entire rapidity range and that the full transparency of the Bjorken
model has not been  reached. 
Nevertheless, reactions at the present
energy evidence the  highest antiparticle/particle ratios so far
observed in energetic nucleus--nucleus collisions.

The BRAHMS collaboration wishes to thank the RHIC team for the great
efforts that have led to the successful startup of the collider and
for the support to the experiment. This work was supported by the
Division of Nuclear Physics of the Office of Science of the
U.S. Department of Energy under contracts DE-AC02-98-CH10886,
DE-FG03-93-ER40773, DE-FG03-96-ER40981, and DE-FG02-99-ER41121, the Danish Natural Science
Research Council, the Research Council of Norway, the Jagiellonian
University Grant, the Korea Research Foundation Grant, and the Romanian
Ministry of Education and Research (5003/1999,6077/2000).

\begin{thebibliography}{99}
\bibitem{Bjorken83} J.~D.~Bjorken, Phys. Rev. {\bf D27}, 140 (1983).
\bibitem{Her99} N.~Herrmann, J.~P.~Wessels and T.~Wienold,
  Ann. Rev. Nucl. Part. Sci. {\bf 49}, 581 (1999).
\bibitem{Vid95} F.~Videb{\ae}k and O. Hansen, Phys. Rev. {\bf C52},
  2584 (1995).
\bibitem{STAR01} C.~Adler et al., The STAR collaboration,
  Phys. Rev. Lett. (in print). 
\bibitem{BRAHMSNIM}
  D.~Beavis et al., `Conceptual Design Report for BRAHMS', BNL-62018;
  I.G. Bearden et al., The BRAHMS collaboration,
  Nucl. Instr.\& Methods, (in preparation).
\bibitem{ZDCNIM} C.~Adler et al., Nucl. Instr. \& Methods,(2001) (in print),
  (/xxx.lanl.gov/nucl-ex/0008005).
\bibitem{SPSppbar} I.~G.~Bearden et al., NA44 collaboration, 
  J. Phys. G, Nucl. Part. {\bf 23}, 1865 (1997); M.~Kaneta Ph. D. Thesis, 1998, University of Hiroshima.
\bibitem{Sick99} F.~Sickler et al., The NA49 collaboration;
  Nucl. Phys. {\bf A661}, 45c (1999); G. E. Cooper, Ph. D. Thesis, 2000, University of California, Berkely.
\bibitem{Ahle98} L.~Ahle et al., The E802 collaboration, Phys. Rev. Lett. {\bf 81}, 2650 (1998).
\bibitem{Guettler76}K.~Guettler et al., Nucl.Phys. {\bf B116}, 77 (1976).
\bibitem{Capi74} P.~Capiluppi et al., Nucl. Phys. {\bf B79}, 189  (1974).
\bibitem{HIJING} HIJING 1.36 with Parton shadowing and Jet
  quenching. X-N.~Wang and  M.~Gyulassy, Phys. Rev. {\bf D44}, 3501 (1991).  
\bibitem{FRITIOF} B.~Anderson et al., Z. Phys. C{\bf 57}, 485 (1993); 
  H.~Pi, Comp. Phys. Comm. {\bf 71}, (1992).
\bibitem{UrQMD} S. A. Bass et al., Prog. Part. Nucl. Phys. {\bf41}
  225 (1998); M. Bleicher et al. J. Phys. G. Nucl. Part {\bf 25}, 1859
  (1999). 
\bibitem{Phobos00} B.~Back et al., Phys. Rev. Lett. {\bf 85}, 3100 (2000).
\end{thebibliography}
\end{document}